\def\half{\frac{1}{2}}
\def\nll{\hfil\hfil\linebreak\noindent} 
\def\dbox#1{\hbox{\vrule  
                        \vbox{\hrule \vskip #1
                             \hbox{\hskip #1
                                 \vbox{\hsize=#1}%
                              \hskip #1}%
                         \vskip #1 \hrule}%
                      \vrule}}
\def\qed{\hfill \dbox{0.05true in}}  
\documentclass{iopart}
\usepackage{graphicx}
\usepackage{cite}
\begin{document}
\hspace*{3.5 in}CUQM-133 \\


\title{Comparison theorems for the Dirac equation with spin-symmetric and pseudo-spin-symmetric interactions}

\markboth{R.~L.~Hall \& \"O. Ye\c{s}ilta\c{s}}{Comparison theorems for the Dirac equation}

\author{Richard~L.~Hall and \"Ozlem Ye\c{s}ilta\c{s}$^{\dagger}$}
\address{Department of Mathematics and Statistics, Concordia
University,\\ 1455 de Maisonneuve Boulevard West, Montr\'eal,
Qu\'ebec, Canada H3G 1M8. \\Email: rhall@mathstat.concordia.ca~and~yesiltas@gazi.edu.tr\\
$^{\dagger}$Permanent address: Department of Physics, Faculty of Arts and Sciences,
Gazi University, 06500 Ankara, Turkey.}

\begin{abstract}
A single Dirac particle is  bound in $d$ dimensions by vector $V(r)$ and scalar $S(r)$ central potentials.
The spin-symmetric $S=V$ and pseudo-spin-symmetric $S =  - V$ cases are studied and it is shown that if
two such potentials are ordered $V^{(1)} \le V^{(2)},$ then corresponding discrete eigenvalues are all
similarly ordered $E_{\kappa \nu}^{(1)} \le E_{\kappa \nu}^{(2)}.$  This comparison theorem allows us to
use envelope theory to generate spectral approximations with the aid of known exact solutions, such
 as those for Coulombic, harmonic-oscillator, and Kratzer potentials.
 The example of the log potential $V(r) = v\ln(r)$ is presented.
Since $V(r)$ is a convex transformation of the soluble  Coulomb potential, this leads to a compact analytical
formula for lower-bounds to the discrete spectrum.  The resulting ground-state lower-bound curve $E_{L}(v)$ is compared with an accurate graph found by direct numerical integration.
\end{abstract}


\vskip 0.2in

\maketitle
\section{Introduction}
Recently there has been much interest in exact solutions of the Dirac or Klein-Gordon equations with scalar and vector potentials of equal magnitude \cite{castro,jia,alhaidari,zhang,dong,jia1}. By the term `vector potential' we  mean the time component $V(r)$ of the energy-momentum four-vector; the scalar potential $S(r)$ is a term added to the mass. The spin and pseudo-spin symmetries in nuclear physics \cite{arima,hecht}, which have been observed in the hadron, are used to explain aspects of deformed nuclei. Spin symmetry occurs in the spectrum of a meson with one heavy quark and anti-nucleon bound in a nucleus \cite{gino1}. Pseudo-spin symmetry occurs in the spectrum of certain nuclei \cite{gino2}. Ginocchio showed that the Dirac Hamiltonian with scalar and vector harmonic oscillator potentials admits both spin-symmetry and $U(3)$ symmetry when $S  = V$, it also admits pseudo-spin symmetry and pseudo-$U(3)$ symmetry when $S = -V$ in the three-dimensional $(d=3)$ case \cite{Ginocchio}. Exact solutions of Dirac equation with a Coulomb-like tensor potential is discussed in \cite{akcay}, under spin and pseudo-spin conditions.

For non-relativistic problems where the Hamiltonian is bounded below and its discrete spectrum can be characterized variationally, the derivation of a comparison theorem of the form
\[
V^{(1)} \le V^{(2)} \Rightarrow E^{(1)} \le E^{(2)}
\]
is almost immediate. The situation is not so clear {\it a priori} for relativistic problems where the corresponding energy operator is not bounded below.  In spite of this difficulty, it has been shown \cite{halld,chen,hallds,hallkg,hallrct} that relativistic comparison theorems are indeed possible, at least with respect to vector potentials, that is to say, the time component $V$ of a four-vector. To our knowledge, no such results have been obtained for comparisons involving a scalar potential $S$ (a variable term added to the mass).  In the present paper we derive comparison theorems for
the Dirac equation in the case of spin-symmetric problems $S  = V$ and for pseudo-spin-symmetric problems $S = -V$.

We consider a single particle that is bound by an attractive central vector and scalar potentials,
respectively $V$ and $S$, in $d\ge 1$ spatial dimensions and obeys the Dirac equation.  For central potentials in $d$ dimensions the Dirac equation can be written \cite{jiang} in natural units $\hbar=c=1$ as
\begin{equation}\label{eq1}
i{{\partial \Psi}\over{\partial t}} =H\Psi,\quad {\rm where}\quad  H=\sum_{s=1}^{d}{\alpha_{s}p_{s}} + (m+S)\beta+V,
\end{equation}
$m$ is the mass of the particle, $V$ and $S$ are the spherically symmetric vector and scalar potentials, and $\{\alpha_{s}\}$ and $\beta$  are Dirac matrices, which satisfy anti-commutation relations; the identity matrix is implied after the vector potential $V$. For stationary states, algebraic calculations in a suitable basis lead to a pair of first-order linear differential equations in two radial functions $\{\psi_1(r), \psi_2(r)\}$, where $r = ||\mathbf{r}||.$  For $d > 1,$ these functions vanish at $r = 0$, and, for bound states, they may be normalized by the relation
\begin{equation}\label{eq2}
(\psi_1,\psi_1) + (\psi_2,\psi_2) = \int\limits_0^{\infty}(\psi_1^2(r) + \psi_2^2(r))dr = 1.
\end{equation}
We use inner products {\it without} the radial measure factor $r^{(d-1)}$ because the factor $r^{\frac{(d-1)}{2}}$ is already built in to each radial function. Thus the radial functions vanish at $r = 0$ and satisfy the coupled equations
\begin{eqnarray}
E\psi_1 &=& (V+m+S)\psi_1 + (-\partial + k_{d}/r)\psi_2\label{eq3}\\
E\psi_2 &=& (\partial + k_{d}/r)\psi_1 + (V-m-S)\psi_2\label{eq4},
\end{eqnarray}
where $k_1 = 0,$ $k_{d}=\tau(j+{{d-2}\over{2}}),~d >1$,  $\tau = \pm 1,$  and the symbol $\partial$ represents the operator $\partial/\partial r.$ We note that the variable $\tau$ is sometimes written $\omega$, as, for example in the book by Messiah \cite{messiah}, and the radial functions are often written $\psi_1 = G$ and $\psi_2 = F,$ as in the book by Greiner \cite{greiner}.  We shall assume that the potentials $V$ and $S$ are such that there are some discrete eigenvalues $E_{k_d \nu}$ and that Eqs.(\ref{eq3},~\ref{eq4}) are the eigenequations for the corresponding radial eigenstates. Here $\nu$ is the number of nodes in the radial wave function for a given $k_d$.  In this paper we shall present the problem explicitly for the cases $d > 1.$ Similar arguments go through for the case $d=1$: in this case $k_1 = 0,$ the states can be classified as even or odd, and the normalization (\ref{eq2}) becomes instead $\int_{-\infty}^{\infty}\left(\psi_1^2(x) + \psi_2^2(x)\right)dx = 1.$

The principal concern of the present paper is with two special cases, namely spin-symmetric problems, for which $S = V$, and pseudo-spin-symmetric problems, for which $S = -V.$ In each class of problems, therefore, there is just one potential function, $V$: it is with respect to this potential that we shall derive comparison  theorems.  We shall first treat the two cases separately,
and then, for the purpose of determining the discrete spectrum, put them together in a single formulation.
\subsection{Spin-symmetric problems $S = V$}
In this case (\ref{eq3}) and (\ref{eq4}) become
\begin{eqnarray}
\psi'_1 +\frac{k_d}{r}\psi_1 &=& (m+E)\psi_2\label{eq5}\\
\psi'_2-\frac{k_d}{r}\psi_2 &=& (m+2V-E)\psi_1\label{eq6}.
\end{eqnarray}
By differentiation and substitution we obtain the following Schr\"odinger-like equation
\begin{equation}\label{eq7}
-\psi''_1 + \left(\frac{k_d(k_d+1)}{r^2} +2(E+m)V\right)\psi_1 = -( m^2-E^2)\psi_1.
\end{equation}
Although the radial function $\psi_1$, which must be $L^2$ because of (\ref{eq2}), is not separately normalized, the eigenequation (\ref{eq7}) does determine the discrete eigenvalue $E.$  This will be  illustrated explicitly in the next section where we shall discuss some exact solutions.  However, we note that (\ref{eq7}) alone is not sufficient to derive the comparison theorem; for this it is still necessary to use the original Dirac equations.
\subsection{Pseudo-spin-symmetric problems $S = -V$}
In this case (\ref{eq3}) and (\ref{eq4}) become
\begin{eqnarray}
\psi'_1 +\frac{k_d}{r}\psi_1 &=& (m+E-2V)\psi_2\label{eq8}\\
\psi'_2-\frac{k_d}{r}\psi_2 &=& (m-E)\psi_1\label{eq9}.
\end{eqnarray}
Again, by differentiation and substitution we obtain a Schr\"odinger equation,  namely
\begin{equation}\label{eq10}
-\psi''_2 + \left(\frac{k_d(k_d-1)}{r^2} +2(E-m)V\right)\psi_2 = -( m^2-E^2)\psi_2.
\end{equation}
\subsection{Combined eigenequation}
By using the following parametrization
\begin{equation}\label{eq11}
\kappa = sk_d = s\tau\left(j+\frac{d-2}{2}\right),\quad \mu = sm,\quad s= \pm 1,
\end{equation}
we  may write the eigenequation for both cases as
\begin{equation}\label{eq12}
-\psi'' + \left(\frac{\kappa(\kappa +1)}{r^2} +2(E+\mu)V\right)\psi = -( \mu^2-E^2)\psi,
\end{equation}
where $\psi$ is assumed to be square  integrable on $[0,\infty),$ and $E = E_{\kappa\nu}$ is a discrete eigenvalue corresponding to
a radial eigenfunction with $\nu = 0,1,2,\dots$ nodes.
\subsection{Principal result}
The principal result of this paper is the following theorem:
\medskip\nll{\bf Theorem}
\begin{equation}\label{eq13}
V^{(1)} \le V^{(2)} \Rightarrow E_{\kappa \nu}^{(1)} \le E_{\kappa \nu}^{(2)}.
\end{equation}
We shall prove this theorem in section~3. In section~2 below we first derive some exact spectral formulae for harmonic-oscillator,
coulombic, Kratzer potentials, and log potentials.  In section~4 we outline the envelope method which provides energy bounds whenever a comparison theorem is available and the given potential can be written as a transformation $V(r) = g(h(r))$ of a soluble potential $h(r),$  where the transformation function $g(h)$ has definite convexity.  In section~5 we consider the log potential $V(r) = v\ln(r)$ which is a convex
transformation of the Coulomb potential $h(r) = -1/r$. We show that envelope theory can use the exact Coulomb solution and the comparison  theorem to generate an implicit analytical formula expressing the dependence of each eigenvalue on the coupling parameter $v.$ The lower-bound curve $E_{L}(v)$ is compared to an accurate graph obtained by scaling and direct numerical integration.
\section{Some exact eigenvalues}
We can obtain exact solutions by comparing the general Dirac eigenequation (\ref{eq12})
for the classes of problems considered with the following generic radial Schr\"odinger equation
\begin{equation}\label{eqs}
-\psi'' +\left(\frac{L(L+1)}{r^2} + vf(r)\right)\psi = {\mathcal E}\psi = F_{\nu L}(v)\psi,
\end{equation}
where $v > 0$ is the coupling parameter for an attractive central potential with shape $f(r),$
$L$ is a generalized angular momentum quantum number that is not necessarily integral, and $F_{\nu L}(v)$
describes how the eigenvalue corresponding to a radial eigenfunction with $\nu$ nodes depends on the coupling.
From the relation $\kappa(\kappa +1) = L(L+1)$ we extract the following formula for $L$:
\begin{equation}\label{eqL}
L  = \left|\kappa +\half\right| -\half = \left|s\tau\left(j+\frac{d-2}{2}\right) +\half\right| -\half,\quad s = \pm 1.
\end{equation}
We now present six illustrations.
\subsection{Harmonic oscillator:~~$V(r) = v r^2$}
From Eq.(\ref{eqs}) in this case we  have
\begin{equation}\label{eqsho}
{\mathcal E} = F_{\nu L}(v) = P\,v^{\half},\quad P = \left(4\nu + 2L + 3\right).
\end{equation}
If we apply this to the eigenequation (\ref{eq12}) for the corresponding Dirac problem,
we obtain
\begin{equation}\label{eqdho}
E^2-\mu^2 = \left(2v(\mu+E)\right)^{\half}P, \quad \mu = sm = \pm m.
\end{equation}
Thus we must have $v(E+\mu) > 0,$ and we conclude $|E| > m$. Numerical values are easily
obtained as the solutions to
\begin{equation}\label{eqdhob}
(E^2-m^2)|E-\mu| = 4P^2|v|.
\end{equation}
satisfying $|E| > m.$
\subsection{The linear potential:~~$V(r) = vr$}
For the linear potential we have by Eq.(\ref{eqs}) and a scaling argument that
\begin{equation}\label{eqslin}
{\mathcal E} = F_{\nu L}(v) = P\,v^{\frac{2}{3}},
\end{equation}
where the Schr\"odinger eigenvalues $P =F_{\nu L}(1)$ for unit coupling $v = 1$ must be determined numerically; these are all positive.
For example, if $d = 3$, $\tau = s = 1$, and $j = \half,$ then from Eq.(\ref{eqL}), $L = 1:$ thus $P = F_{0 1}(1)$ is the bottom of the spectrum of $-\partial ^2 + 2/r^2 + r$, that is to say, $P\approx 3.3612545$.
If we apply this to the eigenequation (\ref{eq12}) for the corresponding Dirac problem,
we obtain
\begin{equation}\label{eqdlin}
E^2-\mu^2 = P\,\left(2v(\mu+E)\right)^{\frac{2}{3}}, \quad \mu = sm = \pm m.
\end{equation}
Thus we must have $v(E+\mu) > 0,$ and we conclude $|E| > m$. Numerical values for Dirac energy $E$ are then
given by solutions to
\begin{equation}\label{eqdlinb}
(E^2-m^2)(E-\mu)^2 = 4v^2 P^3.
\end{equation}
satisfying $|E| > m.$

\subsection{Coulomb potential $V(r) = -v/r$.}
We proceed in a similar way as for the oscillator problem.  In the Coulomb case we have
\begin{equation}\label{eqcoula}
{\mathcal E} = F_{\nu L}(v) = -\frac{v^2}{4P^2},\quad P = \nu + 1 + L.
\end{equation}
If we apply this to the eigenequation (\ref{eq12}) for the corresponding Dirac problem, we obtain
\begin{equation}\label{eqcoulb}
E^2-\mu^2 = -\frac{v^2(\mu+E)^2}{P^2}.
\end{equation}
It is immediately clear that $E^2 < \mu^2 = m^2,$ that is to say, the discrete eigenvalues satisfy $-m < E< m.$  Moreover, we conclude that for spin-symmetric problems ($\mu = m$), we have $v >0$, but for pseudo-spin-symmetric problems ($\mu = -m$), the coupling must be negative, $v < 0.$  We note for future reference that in both Coulomb cases
\begin{equation}\label{eqcoulc}
v\mu > 0.
\end{equation}
For the pure Coulomb problem we have the following explicit spectral formula
\begin{equation}\label{eqcould}
E = \mu\frac{\left(1-\frac{v^2}{P^2}\right)}{\left(1+\frac{v^2}{P^2}\right)}.
\end{equation}
For the Coulomb potential we now briefly discuss the behaviour of $\psi_1$ and $\psi_2$, as solutions of (\ref{eq7}) and (\ref{eq10}), near the origin. The solutions of the combined equation (\ref{eq12}) are given 
\begin{equation}\label{s-wave}
    \psi(r) = c\,r^{L+1} e^{-\sqrt{\mu^{2}-E^{2}}\,r} \mathcal{L}^{2L+1}_{\nu}\left(2\sqrt{\mu^{2}-E^{2}}\,r\right),
\end{equation}
where $c$ is a normalization constant, $\mathcal{L}^{a}_\nu(x)$ are Laguerre polynomials,  and the parameter $L$, which satisfies $L(L+1) = \kappa(\kappa+1)$, is given explicitly by Eq.~(\ref{eqL}).  The small-$r$ asymptotic behaviour of the wave function for the corresponding semirelativistic problem is discussed in Ref. \cite{durand}.
\subsection{Shifted Coulomb potential:~~$V(r) = -v/r + c$}
We now add a constant term $c$ and re-solve, obtaining, first from Eq.~(\ref{eq12}),
\begin{equation}\label{eqcoule}
E^2-\mu^2 - 2c(\mu+E) = -\frac{v^2(\mu+E)^2}{P^2},
\end{equation}
where $P = \nu + L + 1.$  Thus we have for the shifted Coulomb problem
\begin{equation}\label{eqcoulf}
E = \mu\frac{\left(1-\frac{v^2}{P^2}\right)}{\left(1+\frac{v^2}{P^2}\right)} + \frac{2c}{\left(1+\frac{v^2}{P^2}\right)} = -\mu +\frac{2(\mu + c)}{\left(1+\frac{v^2}{P^2}\right)}.
\end{equation}
We note that this result is consistent with Eq.~(52) of Ref.\cite{AS09}. From Eq.~(\ref{eq12}) in this case, it is clear
that, for the existence of discrete eigenvalues, we must have $v(E+\mu) > 0.$ From this inequality and Eq.~(\ref{eqcoulf}) we conclude
\begin{equation}\label{eqcoulcc}
v(c+\mu) > 0.
\end{equation}
Thus, unlike for the Dirac equation with a Coulombic vector potential and a constant scalar potential, where the Coulomb coupling must  not be too large (the $Z < 137$ rule for atoms), here the magnitude $|v|$ of the Coulomb coupling may be chosen as large as we please.  Moreover, unlike for any Schr\"odinger problem, where an added potential constant term $c$ may have any desired value, here $c$  is restricted by Eq.~(\ref{eqcoulcc}).
\subsection{The Kratzer potential:~~$V(r) = a/r^2 -v/r +c$}
The Kratzer  potential \cite{Kratzer, Landau, Fluegge, Morales, Hall98, Znojil, Hooydonk} comprises a
shifted Coulomb potential with an added centrifugal term.  The corresponding eigenvalues are given
 implicitly by the same fomula (\ref{eqcoule}) that we derived above, but now the parameter $P$ depends on $E$.
The new effective $L$ parameter is determined by
\[
\kappa(\kappa +1) + 2a(E+\mu) = L(L+1).
\]
Since $P = \nu+1+L$, we can obtain (for $a\ne 0$) a second expression for $E$ given by
\begin{equation}\label{eqcoulg}
E = \frac{1}{2a}\left[(P-\half-\nu)^2 - (\kappa+\half)^2\right] - \mu.
\end{equation}
Thus, by equating the formulae (\ref{eqcoulf}) and (\ref{eqcoulg}) we obtain a quartic equation whose solution
yields the value of $P$ which then, in turn,  determines $E$ via (\ref{eqcoulf}) or (\ref{eqcoulg}).
We can also use $P$ from  (\ref{eqcoulf}) which is
\begin{equation}\label{Pi}
    P=v\sqrt{\frac{\mu+E}{2c+\mu-E}}
\end{equation}
to obtain an eigenvalue equation which is equivalent to (\ref{eqcoulg}):
\begin{eqnarray}\label{Energy}
    v(\mu+E) & = & \left(\nu+\frac{1}{2} + \sqrt{(\kappa+\frac{1}{2})^{2} \nonumber +
    2a(\mu+E)}\right)\times \\
 & &
\left(2c(\mu+E)+\mu^{2}-E^{2}\right)^{\half}.
\end{eqnarray}
We note that this result is consistent with Eqs. $(40)$ and $(52)$ of Ref.\cite{AS09}. If $c = 0$, then the limit to Coulomb coupling $v = 0$ yields from (\ref{eqcoulf}) $E = \mu = sm =\pm m,$ whatever value we choose for the centrifugal parameter $a.$ We note that this conclusion clearly contradicts the spectral claims made in \S(5.3.2) of Ref.\cite{AS09}; we find no discrete spectrum if the Coulomb coupling is zero, that is to say when $v= 0.$

\subsection{The log potential: $V(r) = v\,\ln(r)$}
We now let the potential shape be $f(r)   = \ln(r)$ in Eq.~(\ref{eqs}).  If we let the eigenvalue with coupling $v = 1$ and given $\nu$ and $L$  be
$$e(1) = e_{\nu L}(1) = F_{\nu L}(1),$$
 then the general eigenvalue with coupling $v>0$ is given by
\begin{equation}\label{eqlogs}
F_{\nu L}(v) = e(1)v -\half v\ln(v).
\end{equation}
We see this by the following scaling argument. The Schr\"odinger Hamiltonian has the form $H = -\Delta + v\ln(r).$  By scaling the radial variable $r\rightarrow r\sigma,$ where $\sigma >0$ is fixed, we merely describe $H$ in a different way, and we obtain the spectrally equivalent operator $H \rightarrow \frac{1}{\sigma ^2}\Delta +v\ln(r\sigma).$  Thus the eigenequation becomes
\begin{equation}\label{eqscale1}
\left[-\left(\frac{1}{\sigma ^2}\right)\Delta +v\ln(r\sigma)\right]\psi = e(v)\psi,
\end{equation}
equivalently,
\begin{equation}\label{eqscale2}
\left[-\Delta +\sigma^2v\ln(r)\right]\psi = \sigma^2(e(v)-v\ln(\sigma))\psi.
\end{equation}
By choosing the scale so that $\sigma^2 v = 1,$ the eigenvalue on the right-hand side of (\ref{eqscale2}) must equal $e(1).$ This establishes Eq.(\ref{eqlogs}). If we now apply this formula to the Dirac combined eigenequation (\ref{eq12}), we find that the Dirac energy $E = E_{\nu L}$ is given by
the following implicit formula:
\begin{equation}\label{eqlogd}
E = \mu + v\left[2e(1)-\ln(2) - \ln(v(\mu + E))\right].
\end{equation}
For later reference we note that the bottom of the spectrum, for arbitrary $v$ (of the appropriate sign) requires the single Schr\"odinger eigenvalue $e(1) = F_{01}(1) \approx 1.6411353.$ From Eq.~(\ref{eq12}) we see that it is always necessary that $u \equiv v(\mu+E)  > 0.$  This parameter has the upper bound $u < u_1,$ where $u_1$ corresponds to $E = 0$: if $u>u_1$, $E$ becomes complex. We have from Eq.~(\ref{eq12}) with $E = 0 $ for this case:
\begin{equation}\label{equ1}
-m^2 = u_1(2e_{\nu L}-\ln(2)) -u_1\ln(u_1).
\end{equation}
For the ground state with $\mu = m = 1$, $d = 3$, and $v > 0$, we have
$\kappa   = L = 1$, $e_{0 1} \approx 1.6411353,$ and $u_1 = v  \approx 14.28389.$
More generally, depending on the signs of $v$ and $\mu = \pm m,$ we may identify
four spectral regions: these are summarized in Table~1.
\begin{table}\label{tab1}
\caption{The four spectral regions for the log potential $V(r) = v\ln(r)$.\\
When $E = 0,$ $u = v(E+\mu)= v\mu = u_1.$}
\begin{indented}
\item[]\begin{tabular}{@{}*{3}{c}}
\br
& $\mu = m$ & $\mu = -m$ \cr
\mr
$v > 0$ & $-m < E < u_1/v-m$ & $m < E < m+ u_1/v$ \cr
$v < 0$ & $-u_1/|v| -m < E < -m$ & $-u_1/|v| + m < E < m$ \cr
\br
\end{tabular}
\end{indented}
\end{table}
\section{Comparison theorem}
We establish the comparison theorem in two steps, the first establishes a differential result, as we did for pure vector potentials in Refs.~\cite{hallds,hallkg}, and the second extends this to a general comparison theorem, as in Ref.~\cite{hallrct}. We therefore begin by considering spin-symmetric or pseudo-spin-symmetric problems in which $V(r,a)$ depends on a parameter $a$ and it is supposed that $\partial V/\partial a \ge 0.$  We shall prove below that this assumption implies $E'(a) \ge 0$ for each discrete eigenvalue.  If we now define the one-parameter family of potentials by
\begin{equation}\label{eqpot}
V(r,a) = V^{(1)}(r) +  a\left(V^{(2)}(r) - V^{(1)}(r)\right),\quad a \in[0,1],
\end{equation}
then $V^{(2)}(r) \ge V^{(1)}(r)$ implies  that $V(r,a)$ is monotone increasing in the parameter $a$.
Thus $E'(a) \ge 0$ implies $E^{(2)}\equiv E(1) \ge E(0)\equiv E^{(1)},$ which result proves the theorem.
It remains to prove the  monotonicity of $E(a).$
\subsection{Proof that $\partial V/\partial a \ge 0 \Rightarrow E'(a) \ge 0.$}
If the normalization integral (\ref{eq2}) is differentiated partially with respect to $a$, and we denote $\partial\psi / \partial a=\psi_{a}$ for each wave-function component, we obtain the orthogonality relation
\begin{equation}\label{ortho}
    (\psi_{1a},\psi_{1})+(\psi_{2a},\psi_{2})=0.
\end{equation}
Now we differentiate (\ref{eq3}) and (\ref{eq4}) with respect to $a$ to obtain
\begin{eqnarray}\label{deriv1}
    E'(a) \psi_{1}+E(a) \psi_{1a} & = & (V_{a}+S_{a})\psi_{1}+(V+m+S)\psi_{1a} \\ & &+(-\partial+k_{d}/r)\psi_{2a}, \nonumber
\end{eqnarray}
and
\begin{eqnarray}\label{deriv2}
    E'(a) \psi_{2}+E(a) \psi_{2a} & = & (V_{a}-S_{a})\psi_{2}+(V-m-S)\psi_{2a} \\ & & +(\partial+k_{d}/r)\psi_{1a}. \nonumber
\end{eqnarray}
The linear combination $(\ref{deriv1})\psi_{1}+(\ref{deriv2})\psi_{2}$ of these two equations may be written
\begin{eqnarray}\label{g}
    E'(a) [(\psi_{1},\psi_{1})+(\psi_{2},\psi_{2})] & = & (\psi_{1}, (S_{a}+V_{a})\psi_{1}) \\ & & +(\psi_{2}, (V_{a}-S_{a})\psi_{2})+W \nonumber
\end{eqnarray}
where
\begin{eqnarray}\label{W}
    W & = & \left(\psi_{2},(\partial+k_{d}/r) \psi_{1a}\right)-\left(\psi_{1},(\partial-k_{d}/r)\psi_{2a}\right) \\ \nonumber
& & -  \left((m+E(a)+S-V)\psi_{2a},\psi_{2}\right)\\ \nonumber
& & +\left((m-E(a)+S+V)\psi_{1a},\psi_{1}\right).
\end{eqnarray}
We have again used $\partial$ to denote the differential operator $\partial=\partial/\partial r$.  The boundary conditions imply the antisymmetric relation
\begin{equation}\label{anti}
    (\psi, \partial\phi)=-(\partial\psi, \phi).
\end{equation}
If (\ref{anti}) is used in (\ref{W}), it becomes
\begin{eqnarray}
\label{W1}
    W & = & (\psi_{1a},[(m-E(a)+S+V)\psi_{1}+(-\partial+k_{d}/r)\psi_{2}]) \\ \nonumber
& & +(\psi_{2a},[-(m+E(a)+S-V)\psi_{2}-(-\partial-k_{d}/r)\psi_{1}]).
    \end{eqnarray}
From equations (\ref{eq3}) and (\ref{eq4}) it is clear that $W = 0.$ Thus the expression for $E'(a)$ becomes
in the spin-symmetric case $S = V$
\begin{equation}\label{Ea}
    E'(a)[(\psi_{1},\psi_{1})+(\psi_{2},\psi_{2})]= 2\left(\psi_{1},\frac{\partial V}{\partial a}\psi_{1}\right)
\end{equation}
and in the pseudo-spin-symmetric case $S = -V$
\begin{equation}\label{Eap}
    E'(a)[(\psi_{1},\psi_{1})+(\psi_{2},\psi_{2})]= 2\left(\psi_{2}, \frac{\partial V}{\partial a}\psi_{2}\right).
\end{equation}
Thus the theorem is established since, in either case, if $\partial V/\partial a$ has a definite sign, then $E'(a)$ necessarily has the same sign.\qed

The Coulomb problem of \S 2.2 provides an illustration if we let $a = v.$
We have for the shifted Coulomb potential $V(r) = -v/r   + c$ that $\partial V/\partial v = -1/r < 0.$ Meanwhile, we obtain from the corresponding spectral formula (\ref{eqcoule})
\begin{equation}\label{eqcoulep}
E'(v) = - \frac{4(\mu+c)v}{P^2(1+v^2/P^2)^2}.
\end{equation}
Since, from Eq.~(\ref{eqcoulcc}), $(\mu +c)v >0$ for the shifted Coulomb problem, we conclude $E'(v) <0,$ as  predicted by the comparison theorem.

\section{Envelope theory}
Envelope theory \cite{env1,env2,env3,env4} is based on a simple geometrical idea.  If the potential $V(r)$ may be written as a smooth transformation $V(r) = g(h(r))$ of a soluble potential $h(r)$, and the transformation function $g$ has definite convexity, then the tangents to $g(h)$, all of the form $bh(r) +c$, lie entirely above or below $V(r)$. If the eigenvalue problems satisfy a comparison theorem, then these shifted $h$-potentials provide upper or lower spectral bounds.  The envelope method consists of finding the best energy bound from one of these families. Suppose, for example, that $g(h)$ is a convex function ($g''(h) \ge 0$), then in this case we obtain a family of lower potentials given by
\begin{equation}\label{eqtpot}
V(r) \ge V^{(t)}(r) = b(t)h(r) + c(t),
\end{equation}
where
\begin{equation}\label{eqab}
b(t) = g'(h(t)),\quad c(t) = g(h(t)) - h(t)g'(h(t)),
\end{equation}
and $r = t$ is the point of contact between the tangent $bh(r) + c$ and the potential $V(r)$.
\section{The log potential}
We now consider the log potential $V(r) = v\ln(r)$, where $v$ is a coupling parameter: $v$ is positive in the spin-symmetric case and negative in the pseudo-spin-symmetric case. For a soluble envelope basis we choose the Coulomb potential $h(r) = -1/r.$  Thus for the log potential shape we have
\begin{equation}\label{eqlogpotg}
f(r) = \ln(r) = g(h(r)) = -\ln(-h(r)).
\end{equation}
It follows that $g'(h) = -1/h>0$ and $g''(h) = 1/h^2>0$, that is to say, $g$ is monotone increasing and convex.
Equation (\ref{eqab}) then provides us with the $b(t)$ and $c(t)$ coefficients: $b(t) = t$ and $c(t) = 1+\ln(t).$   The potential-shape inequality in this case therefore becomes
\begin{equation}\label{ineq}
    \ln(r) = f(r)\geq f^{(t)}(r)= -\frac{t}{r} + (1+\ln(t)).
\end{equation}
In view of the comparison theorem and the exact eigenvalue formula (\ref{eqcoulf}) for the shifted Coulomb potential, we are now able to construct spectral-bounds.  Since the direction of
the full potential inequality, including the coupling,  now depends on the sign of the coupling $v$ parameter: if $v >0,$ we obtain a lower bound; if $v <0,$ envelope theory generates an upper bound.  The bound, once established, may then be optimized over $t$. For our illustration we take $v >0$ for the spin-symmetric case $\mu = m$, and we find
\begin{equation}\label{eqlogLt}
    E_{L} =   \max_{t}\left[\mu\frac{\left(1-\frac{(vt)^2}{P^2}\right)}{\left(1+\frac{(vt)^2}{P^2}\right)} + \frac{2v(1+\ln(t))}{\left(1+\frac{(vt)^2}{P^2}\right)}\right].
\end{equation}
It is clear from this equation that as $v\rightarrow 0$, $E_{L}\rightarrow \mu$, and as $v \rightarrow \infty$, $E_{L}\rightarrow  -\mu.$ By working with a new optimization parameter $q = (vt/P)^2,$ we can re-write Eq(\ref{eqlogLt}) in the form
\begin{equation}\label{eqlogLq}
    E_{L} =   \max_{q}\left[\mu\left(\frac{1-q}{1+q}\right) + \frac{2v(1+\ln(P/v)) +v\ln(q)}{1+q}\right].
\end{equation}
After some algebra, we find that the critical value of $q$ satisfies $q = v/(\mu + E),$ and, moreover, it is possible to obtain the following implicit analytical formula for all the eigenvalues:
\begin{equation}\label{eqlogL}
E_{L} = m + v\left[1 + 2\ln(P) - \ln\left(v(m+E_{L})\right)\right],
\end{equation}
where $P = 1+\nu + L$ and $L$ is given by Eq.~(\ref{eqL}).  In Fig.~1 we exhibit the energy curve $E_{L}(v)$ for the spin-symmetric ground state ($P = 2$) for $\mu = m  = 1$, and in dimension $d=3$, along with a corresponding accurate curve $E(v)$ obtained from Eq.~(\ref{eqlogd}).
\medskip

\begin{figure}[htbp]\centering\includegraphics[width=12cm]{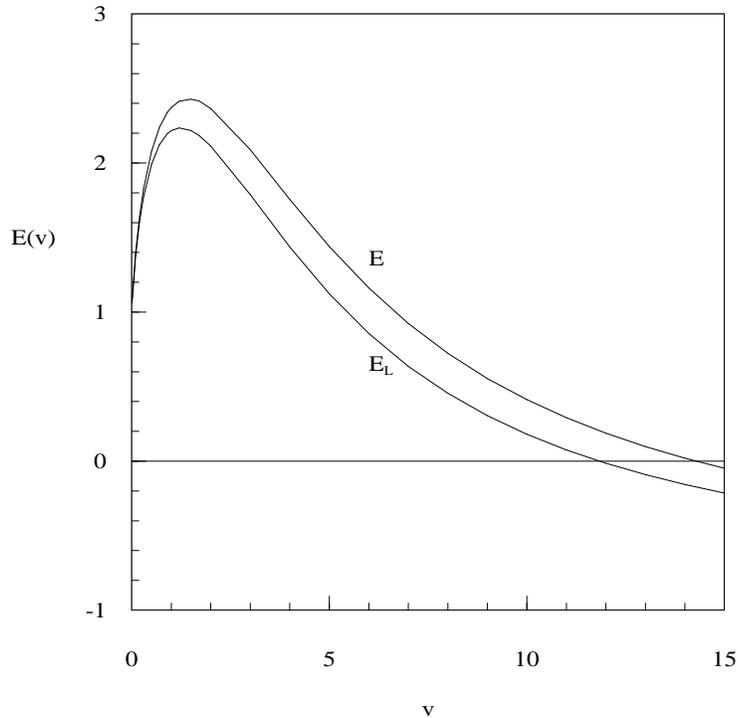}
\caption{Ground-state spin-symmetric energy $E(v)$ for the potential $V(r) = v \ln(r)$, with
 $m = 1$, $d = 3$, and $\kappa = L = 1.$ The figure shows the envelope lower bound $E_{L}$ provided by Eq.~(\ref{eqlogL})
and accurate numerical values $E$ given by Eq~(\ref{eqlogd}) with $e(1) = 1.6411353.$}\label{Fig:E(v)}\end{figure}

\section{Conclusion}
Comparison theorems immediately generate spectral approximations.
 Thus spectral data from one problem yields spectral estimates for another.
Until recently, this kind of reasoning was not expected to be valid for relativistic
systems since their discrete spectra are not easily characterized variationally.
It was then discovered that comparison theorems could be established without
recourse to variational arguments. This  has led to some comparison theorems
for relativistic systems with different vector potentials.  In this paper we
have shown that spin-symmetric and pseudo-spin-symmetric Dirac problems also admit
comparison theorems.  This in turn allows us to use envelope theory to estimate the
spectra of such systems.  For the log potential we are thus able to generate a spectral formulae
which provide energy bounds for each discrete eigenvalue.

\section*{Acknowledgements}

One of us (RLH) gratefully acknowledges partial financial support
of this research under Grant No.\ GP3438 from the Natural Sciences
and Engineering Research Council of Canada; and one of us (\"OY) would like
to thank the Department of Mathematics and Statistics of Concordia University
 for its warm hospitality.

\medskip



\end{document}